\documentclass[conference]{IEEEtran}
\IEEEoverridecommandlockouts
\usepackage{cite}
\usepackage{amsmath,amssymb,amsfonts}
\usepackage{algorithmic}
\usepackage{graphicx}
\usepackage{textcomp}
\usepackage{xcolor}
\usepackage[bookmarks=false]{hyperref}
\usepackage{bm}
\usepackage{fancyhdr}

\def\BibTeX{{\rm B\kern-.05em{\sc i\kern-.025em b}\kern-.08em
    T\kern-.1667em\lower.7ex\hbox{E}\kern-.125emX}}
\begin{document}

\title{DRIFT: Joint Channel Estimation and Prediction Towards Pilotless 6G Non-Terrestrial Networks \\
}

\author{\IEEEauthorblockN{Bruno De Filippo\IEEEauthorrefmark{1}, Carla Amatetti\IEEEauthorrefmark{1}, Alessandro Vanelli-Coralli\IEEEauthorrefmark{1}}
\IEEEauthorblockA{\IEEEauthorrefmark{1}Department of Electrical, Electronic, and Information Engineering (DEI), Univ. of Bologna, Bologna, Italy}
\{bruno.defilippo, carla.amatetti2, alessandro.vanelli\}@unibo.it}

\maketitle
\thispagestyle{fancy}

\begin{abstract} 
Non-terrestrial networks (NTNs) are expected to play a pivotal role in sixth-generation (6G) systems by enabling ubiquitous connectivity and massive communication. In this context, channel prediction emerges as a key technique to improve the spectrum utilization efficiency by limiting the pilot overhead. However, many proposed predictors based on artificial intelligence (AI) are characterized by high inference complexity, posing challenges to onboard implementation. In this paper, we address the challenge of designing accurate yet computationally efficient channel prediction techniques tailored to low Earth orbit (LEO) NTNs, where strict power constraints limit model complexity, to enable spectral efficiency gains. We propose an iterative joint channel estimation and prediction framework in the context of 6G NTNs that significantly reduces pilot overhead by transmitting pilots only in the initial slot and relying on data-driven processing for subsequent slots. We introduce Data-driven Refinement and Iterative Forecast for wireless channel Tracking (DRIFT), a lightweight architecture that refines data-aided channel estimates and predicts future channel frequency responses with low computational cost and reduced error propagation. Two predictor variants based on convolutional and long short-term memory layers are investigated. Simulation results in an end-to-end simulation of an uplink LEO NTN scenario show that the proposed approach achieves up to 12\% spectral efficiency gain compared to conventional pilot-based systems, with robustness to training-test mismatches and consistent performance across different channel models. Moreover, DRIFT requires fewer than 200k multiply-accumulate operations, making it suitable for on-board satellite implementation under stringent power constraints.
\end{abstract}


\section{Introduction}\label{ch:1_intro}
Non-terrestrial networks (NTNs), encompassing satellite, high-altitude platform, and unmanned aerial vehicle communications, are emerging as a cornerstone of future sixth-generation (6G) systems due to their ability to provide ubiquitous connectivity beyond the limits of terrestrial infrastructure \cite{bib:6GNTN}. At the same time, artificial intelligence (AI) is widely recognized as one of the fundamental technologies for the evolution from 5G to 6G, with new base stations being denoted in the 3rd generation partnership project (3GPP) as aNodeBs (aNBs) as a possible reference to the AI-native nature of 6G \cite{bib:aNB}. In particular, technical report 38.843 has identified three main use cases for the application of AI in the radio access network, among which is channel prediction \cite{bib:tr38.843}. This is particularly relevant in the context of low Earth orbit (LEO) NTNs, where the orbit dynamics introduce strong Doppler shifts and delays in a time-varying but deterministic manner. Indeed, several authors have proposed channel prediction solutions to exploit such dynamics with the objective of countering the so-called channel aging effect, \textit{e.g.}, with a long short-term memory (LSTM)-based model in \cite{bib:channelAgingSat} and with a convolutional neural network (CNN) and an LSTM layer in \cite{bib:upToDown}. A transformer-based model was proposed in \cite{bib:TFformer_pred} to predict a time series of channel frequency response (CFR) matrices based on historical data in multi-antenna LEO-based NTNs, achieving state-of-the-art normalized mean squared error (NMSE) figures while requiring more than 100M floating point operations (FLOPs) for inference. In \cite{bib:SCP}, the time series prediction task was carried out with a simpler LSTM-based model. An LSTM-based model was also used in \cite{bib:QV_pred} to predict the excess path loss in Q/V band NTN links. Finally, the authors in \cite{bib:attentionPred} incorporated the attention mechanism in a CNN+LSTM model taking into account not only historical channel data, but also the LEO satellite's elevation angle.

Differently from terrestrial networks, NTNs typically operate under strict power budget constraints, severely limiting the computational complexity of deep learning models. Thus, the design of a channel predictor tailored to NTNs is a tradeoff between prediction accuracy and computational complexity. Nonetheless, most of the works in the literature tend to emphasize to former over the latter. Furthermore, the impact of channel prediction inaccuracies on end-to-end metrics such as the effective spectral efficiency (SE) and throughput is rarely evaluated, with many works focusing on the NMSE instead. In our previous works \cite{bib:commStd_NOMApred, bib:JWCN_pred} we aimed at covering such aspects by proposing lightweight channel predictors based on temporal convolutional networks (TCNs) and CNN-LSTM tailored to NTNs with the objective of reducing the pilot overhead and, thus, improving the throughput by up to $\approx 8.33$\%. In this paper, we build on top of such works by:
\begin{itemize}
    \item Presenting an iterative joint channel estimation and prediction framework where pilot-aided estimation is replaced with channel refinement and prediction;
    \item Proposing the Data-driven Refinement and Iterative Forecast for wireless channel Tracking (DRIFT) model, which refines data-aided (DA) channel estimates and predicts the upcoming CFR with extremely low complexity;
    \item Evaluating the SE gains achieved with respect to traditional estimation-based systems in a simulated LEO-based NTN link, also considering mismatches between training and test data.
\end{itemize}

\section{System model}\label{ch:2_systemModel}
We consider an uplink transmission from a single user equipment (UE) to an aNB hosted on a LEO satellite, as in a 6G NTN scenario. The UE employs cyclic prefix orthogonal frequency division multiplexing (CP-OFDM) and is assumed to pre-compensate the deterministic Doppler and delay components induced by the satellite mobility through a global navigation satellite system (GNSS) receiver, so that only a residual frequency synchronization error affects the link. The transmitter chain includes low-density parity-check (LDPC) channel encoding, bit interleaving, $M$-ary quadrature amplitude modulation (QAM) mapping, and CP-OFDM multiplexing. Let $N_{SC}$ denote the number of subcarriers allocated to the UE, and let $N_{sym}^{slot}$ be the number of OFDM symbols per slot. The transmitted resource grid in slot $t$ is represented as:
\begin{equation}
    \mathbf{X}_t \in \mathbb{C}^{N_{SC}\times N_{sym}^{slot}}.
\end{equation}
We model the time-frequency-selective channel as a tapped-delay line (TDL) with a residual carrier frequency offset $\xi\sim \mathcal{N}(0,\sigma_f^2)$ due to imperfect frequency synchronization, with $\sigma_f^2$ being the variance of the carrier frequency offset. Under the assumption that $\xi$ is small compared to the subcarrier spacing $\Delta f$, the received grid can be obtained as:
\begin{equation}
    \mathbf{Y}_t = \mathbf{H}_t \odot \mathbf{X}_t + \mathbf{W}_t,
\end{equation}
where $\mathbf{H}_t \in \mathbb{C}^{N_{SC}\times N_{sym}^{slot}}$ is the CFR matrix in slot $t$, and $\odot$ denotes the Hadamard product. $\mathbf{W}_t\in \mathbb{C}^{N_{SC}\times N_{sym}^{slot}}$ models additive white Gaussian noise (AWGN), \textit{i.e.}, $\left(\mathbf{W}_t\right)_{k,n}\sim \mathcal{N}(0, N_0/2)$, where $\left(\mathbf{W}_t\right)_{k,n}$ denotes the element at the $k$-th subcarrier and $n$-th OFDM symbol of $\mathbf{W}_t$ and $N_0$ is the receiver's noise power spectral density.
\begin{figure}[t]
    \centerline{\includegraphics[width=0.9\columnwidth]{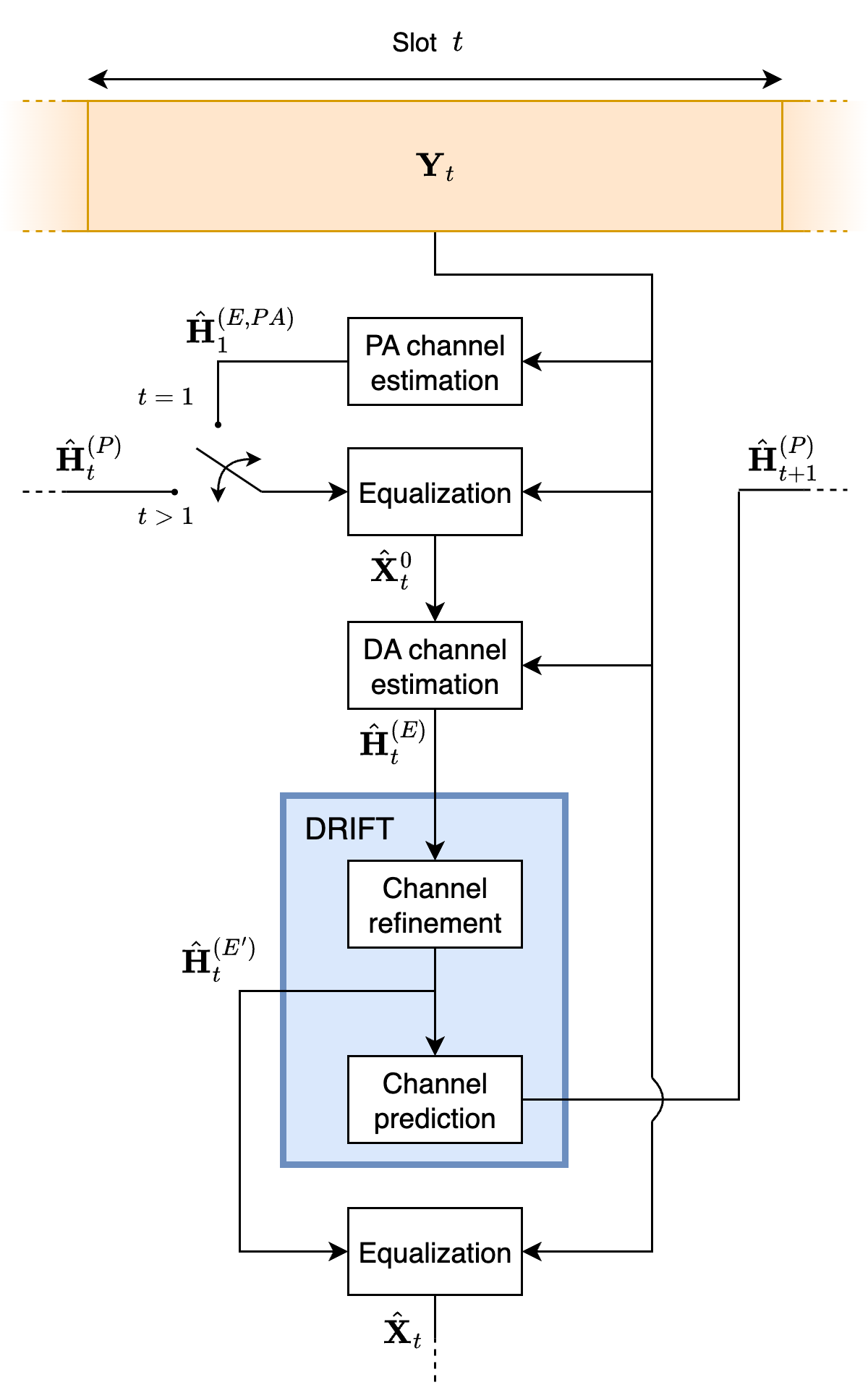}}
    \caption{Iterative joint channel estimation and prediction framework.}
    \label{fig:framework}
\end{figure}
The proposed receiver operates in an iterative prediction loop across $N_{slot}$ consecutive slots. In the first slot, denoted by $t=1$, pilot symbols are inserted in the resource grid to initialize the process as in a traditional estimation-based system. Specifically, we alternate quadrature phase shift keying (QPSK) pilot symbols with null symbols over the subcarriers at OFDM symbol indices $\mathcal{I}_{\pi} = \{3,12\}$ (1-based indexing), therefore boosting the pilot power. All the remaining resource elements of the first slot carry data symbols. For the subsequent slots, $2 \leq t \leq N_{slot}$, no pilots are transmitted and the resource grid carries only data symbols. Therefore, the theoretical error-free SE gain achieved through this framework can be evaluated as the gain in user data resource elements over $N_{slot}$ slots:
\begin{equation}\label{eqn:se_gain}
    \Gamma(N_{slot}) = \frac{N_{slot}N_{sym}^{slot}-\lvert \mathcal{I}_{\pi} \rvert}
    {N_{slot}\left(N_{sym}^{slot} - \lvert \mathcal{I}_{\pi} \rvert\right)}-1.
\end{equation}
It is worth noting that, for an infinite sequence of pilotless slots, such gain approaches $\Gamma_\infty=N_{sym}^{slot}/(N_{sym}^{slot} - \lvert \mathcal{I}_{\pi} \rvert) -1 \approx 16.67$\%. Nonetheless, the formulation in \eqref{eqn:se_gain} quickly leads to diminishing returns, \textit{e.g.}, $\Gamma(5)\approx13.33$\%. Since the first slot is the only one containing pilots, a conventional least-squares (LS) channel estimate is first computed on the pilot positions:
\begin{equation}
    \hat{\mathbf{H}}^{(E,PA)}_{1} = Interp\left[
    \left(\mathbf{X}_{1}\right)_{:,\mathcal{I}_{\pi}}^{-1}\left(\mathbf{Y}_{1}\right)_{:,\mathcal{I}_{\pi}}\right],
\end{equation}
where spline interpolation is used to obtain channel estimates outside of pilot locations. In a traditional system, this pilot-aided (PA) estimate is computed and used for equalization on each slot; on the opposite, our framework employs it to equalize data symbols during the first slot only and to bootstrap the iterative procedure. For a generic slot $t$, the received grid is equalized by means of the channel matrix predicted at the previous iteration, denoted by $\hat{\mathbf{H}}^{(P)}_{t}$, resulting in the equalized grid $\hat{\mathbf{X}}^{0}_{t}$. In the first slot, we simply set $\hat{\mathbf{H}}^{(P)}_{1}=\hat{\mathbf{H}}^{(E,PA)}_{1}$. Hard decisions are taken on $\hat{\mathbf{X}}^{0}_{t}$ by collapsing each received symbol to the closest $M$-QAM constellation point. Then, a DA channel estimation stage performs LS channel estimation on $\mathbf{Y}_t$ considering such reconstructed symbols as ground truth, obtaining the DA channel estimate $\hat{\mathbf{H}}^{(E)}_{t}$. Through DA channel estimation, we inject updated information from the current time slot in the iterative process. Clearly, wrong decisions will affect the accuracy of $\hat{\mathbf{H}}^{(E)}_{t}$: the DRIFT architecture aims at equalizing such errors and produce a refined channel estimate $\hat{\mathbf{H}}^{(E')}_{t}$ to be used for equalization in pilotless slots.

\begin{figure*}[t]
    \centerline{\includegraphics[width=\textwidth]{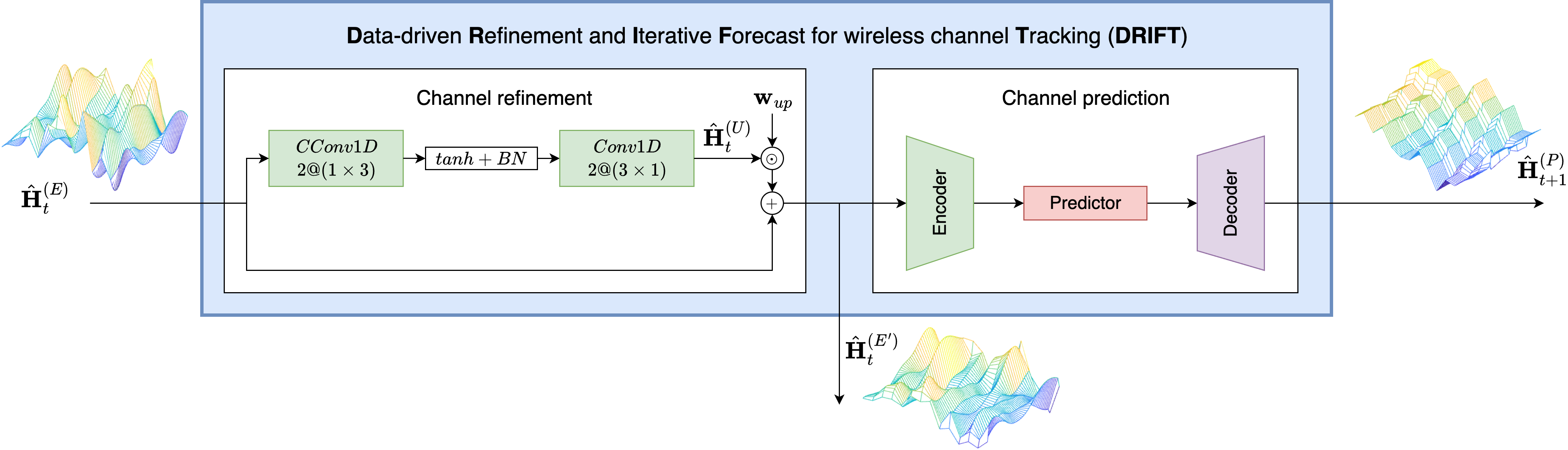}}
    \caption{Diagram of the DRIFT model.}
    \label{fig:CNNscheme}
\end{figure*}

\section{Joint channel refinement and prediction}\label{sec:3_CNN}
The proposed DRIFT architecture, reported in Figure \ref{fig:CNNscheme}, comprises two sections: channel refinement and channel prediction. Both sections operate on three-dimensional real-valued tensors (not accounting for the batch dimension) obtained by stacking the real and imaginary part of the corresponding unit-norm CFR over the third axis. For the sake of simplicity, we maintain the previously introduced CFR terminology (\textit{e.g.}, $\hat{\mathbf{H}}_{t}^{(E)}$) to refer to the three-dimensional tensor of each CFR in this Section.

\subsection{Channel refinement}
At time slot $t$, this block takes as input the DA-estimated CFR $\hat{\mathbf{H}}_{t}^{(E)}$ and aims at equalizing the errors introduced through inaccurate demapping, thus producing a more accurate version $\hat{\mathbf{H}}_{t}^{(E')}$. A CFR update $\hat{\mathbf{H}}_{t}^{(U)}$ is generated through a compact update sequence encompassing: 1) a temporal-only one-dimensional causal convolution (CConv1D) layer; 2) a hyperbolic tangent activation function; 3) batch normalization (BN); and 4) a spectral-only (\textit{i.e.}, operating over the subcarriers only) one-dimensional convolutional (Conv1D) layer. We introduce an update weight vector $\mathbf{w}_{up}\in\mathbb{R}^{N_{sym}^{slot}\times 1}$ to scale the CFR update before applying it to $\hat{\mathbf{H}}_{t}^{(E)}$ by means of summation. As the considered framework foresees $\hat{\mathbf{H}}_{t}^{(E)}$ being based on $\hat{\mathbf{H}}_{t}^{(P)}$, resulting in the estimate's accuracy decreasing over time, we emphasize updates on unreliable (\textit{i.e.}, further in time) estimates through the following formulation:
\begin{equation}\label{eqn:weight}
    (\mathbf{w}_{up})_n = \frac{n}{\frac{1}{N_{sym}^{slot}}\sum_{m=1}^{N_{sym}^{slot}}m},
\end{equation}
where $(\mathbf{w}_{up})_n$ denotes the $n$-th element of $\mathbf{w}_{up}$. This results in a unit mean vector of linearly increasing weights. The refined CFR can then be obtained as follows:
\begin{equation}
    \hat{\mathbf{H}}_{t}^{(E')} =  \hat{\mathbf{H}}_{t}^{(E)} + \mathbf{w}_{up} \odot \hat{\mathbf{H}}_{t}^{(U)},
\end{equation}
where $\mathbf{w}_{up}$ is broadcasted to the subcarriers and real/imag dimensions.

\subsection{Channel prediction}
The objective of the channel prediction block in Figure \ref{fig:CNNscheme} is to obtain at time slot $t$ an estimate for the CFR matrix expected at the subsequent time slot, $\mathbf{H}_{t+1}$, based on the refined CFR $\hat{\mathbf{H}}_{t}^{(E')}$. To maintain computational complexity at a minimum, the model's structure is generally divided into encoder, predictor, and decoder. We here implement the following two channel prediction models:
\begin{itemize}
    \item \textbf{Hybrid CNN-LSTM} \cite{bib:JWCN_pred}. The model employs Conv2D and transposed Conv2D (TConv2D) layers to compress the input CFR $\hat{\mathbf{H}}_{t}^{(E')}$ and reconstruct the output CFR $\hat{\mathbf{H}}_{t+1}^{(P)}$, respectively. The predictor is based on an LSTM layer operating on the compressed CFR representation. To enhance its prediction capabilities, the compressed CFR representation is mirrored on the temporal axis before being fed to the LSTM layer. Further details and parameters are reported in \cite{bib:JWCN_pred}.
    \item \textbf{TCN} \cite{bib:commStd_NOMApred}. The TCN approaches the prediction task through feed-forward temporal convolutional layers, thus avoiding recurrent structures. The predictor stage leverages residual blocks with dilated Conv1D layers operating over the temporal axis. In parallel, skip connections with additional convolutional layers retain multi-scale temporal features from earlier stages, enriching the learned representation. The encoder and decoder stages are based on Conv2D layers, with compression and expansion being carried out through average pooling/unpooling. Further details and parameters are reported in \cite{bib:commStd_NOMApred}.
\end{itemize}

\subsection{Dataset, training, and inference}\label{ch:3.3_DataTrainInfer}
\begin{figure*}[h]
    \normalsize
    \begin{equation}\label{eqn:NMSE}
        \mathcal{L} = \frac{1}{N_{sym}^{slot}N_{SC}}\left(
        \gamma^{(E)}\sum_{n=1}^{N_{sym}^{slot}}\sum_{k=1}^{N_{SC}}
        \frac{\lVert (\hat{\mathbf{H}}_{1}^{(E')})_{k,n} - (\mathbf{H}_{1})_{k,n}\rVert^2}{\lVert (\mathbf{H}_{1})_{k,n} \rVert^2} + 
        (1-\gamma^{(E)})\sum_{n=1}^{N_{sym}^{slot}}\sum_{k=1}^{N_{SC}}
        \frac{\lVert (\hat{\mathbf{H}}_{2}^{(P)})_{k,n} - (\mathbf{H}_{2})_{k,n}\rVert^2}{\lVert (\mathbf{H}_{2})_{k,n} \rVert^2}
        \right)
    \end{equation}
    \hrulefill
\end{figure*}
At the beginning of each training epoch, we generate a synthetic dataset containing $N_{B}$ data samples based on the system model described in Section \ref{ch:2_systemModel} setting $N_{slot}=2$. Each sample point includes: 1) the channel matrix estimated at the first time slot, $\hat{\mathbf{H}}_{1}^{(E)}$; 2) the corresponding ground truth CFR $\mathbf{H}_{1}$; and 3) the ground truth CFR at the subsequent time slot $\mathbf{H}_{2}$. The training loss is the normalized mean squared error (NMSE) computed over both outputs, reported in \eqref{eqn:NMSE} at the top of the current page (the batch dimension is not included in the formulation for the sake of clarity). In a curriculum learning fashion, we initialize the weight $\gamma^{(E)}$ to $\gamma^{(E)}_{max}=1$ at the beginning of training and linearly scale it down to $\gamma^{(E)}_{min}=0.7$ over the span of 1000 epochs, after which it remains constant. In doing so, the DRIFT model initially learns to accurately refine $\hat{\mathbf{H}}_{1}^{(E)}$ and progressively moves towards prediction. The Adam optimizer is employed together with learning rate warm-up and cosine annealing with warm restarts. Training is interrupted once 3 cosine annealing cycles have been completed without a loss improvement. As online learning is outside of the scope of this work, the pre-trained DRIFT model can be directly implemented at the satellite-side receiver for inference.

\section{Results}\label{ch:4_results}
\setlength{\tabcolsep}{5pt}
\renewcommand{\arraystretch}{1.1}
\begin{table}[t]
    \centering
    \caption{Simulation parameters}
    \label{tab:parameters}
    \begin{tabular}{|c|c|}
        \hline
        \textbf{Parameter} & \textbf{Value} \\
        \hline
        Satellite altitude & 600 km \\
        \hline
        Training $E_b/N_0$ range & $E_b/N_0 = [0, 1, ..., 10]$ [dB] \\
        \hline
        Maximum Monte Carlo Iterations & $N_{MC} = 10^5$ \\
        \hline
        Data modulation order (training) & $M^{(training)}=16$ \\
        \hline
        Data modulation order (test) & $M=4$ \\
        \hline
        Code rate & $R_c = 3/4$ \\
        \hline
        Fading model & \{NTN-TDL-C, NTN-TDL-D\} \cite{bib:tr38.811} \\
        \hline
        UE speed & $5$ km/h \\
        \hline
        Carrier frequency & $f_c = 2$ GHz\\
        \hline
        Delay spread & $30$ ns\\
        \hline
        Number of subcarriers & $N_{SC} = 48$\\
        \hline
        Number of slots & $N_{slot} \in \{1, ..., 5\}$\\
        \hline
        Subcarrier spacing & $\Delta f = 15$ kHz\\
        \hline
        Pilot indices per slot & $\mathcal{I}_\pi = \left[3, 12\right]$\\
        \hline
        Batch size & $N_B = 2048$\\
        \hline
        Learning rate range & $[10^{-4}, 6\cdot 10^{-3}]$\\
        \hline
        Learning rate warm-up duration & $40$ epochs\\
        \hline
        Cosine annealing period & $100$ epochs\\
        \hline
        Early stopping patience & $3$ cosine annealing cycles\\
        \hline
    \end{tabular}
\end{table}
\begin{figure}[t]
    \centerline{\includegraphics[width=\columnwidth]{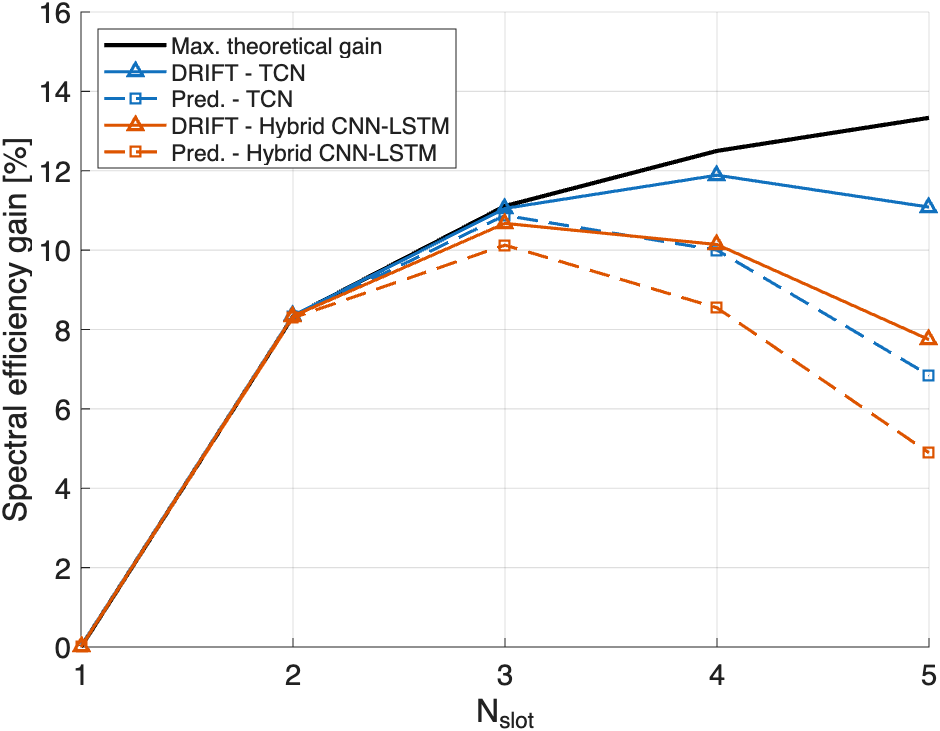}}
    \caption{SE gain as a function of $N_{slot}$ with DRIFT and simple prediction at $E_b/N_0 = 10$ dB (NTN-TDL-C).}
    \label{fig:se_gain}
\end{figure}
We assess the performance of DRIFT models in an end-to-end simulation in the MATLAB computing environment, where a new resource grid spanning $N_{slot}$ slots (each containing $N_{sym}^{slot}=14$ OFDM symbols) is generated at each Monte Carlo iteration. The simulation parameters are reported in Table \ref{tab:parameters}. To simulate a frequency synchronization error with 0.1 parts per million of $f_c$ as per 5G requirements \cite{bib:sync}, we set $\sigma_D = \frac{0.1\cdot 10^{-6}\cdot f_c}{3}$ based on the $3\sigma$ rule. We train the DRIFT models and the corresponding standalone predictors with 16-QAM data symbols over the NTN-TDL-C line of sight (LoS) fading model \cite{bib:tr38.811}. On the opposite, all tests are carried out considering QPSK data symbols, thus introducing a misalignment between training and test data.

\subsection{Spectral efficiency}\label{ch:4.1_SE}
\begin{figure}[t]
    \centerline{\includegraphics[width=\columnwidth]{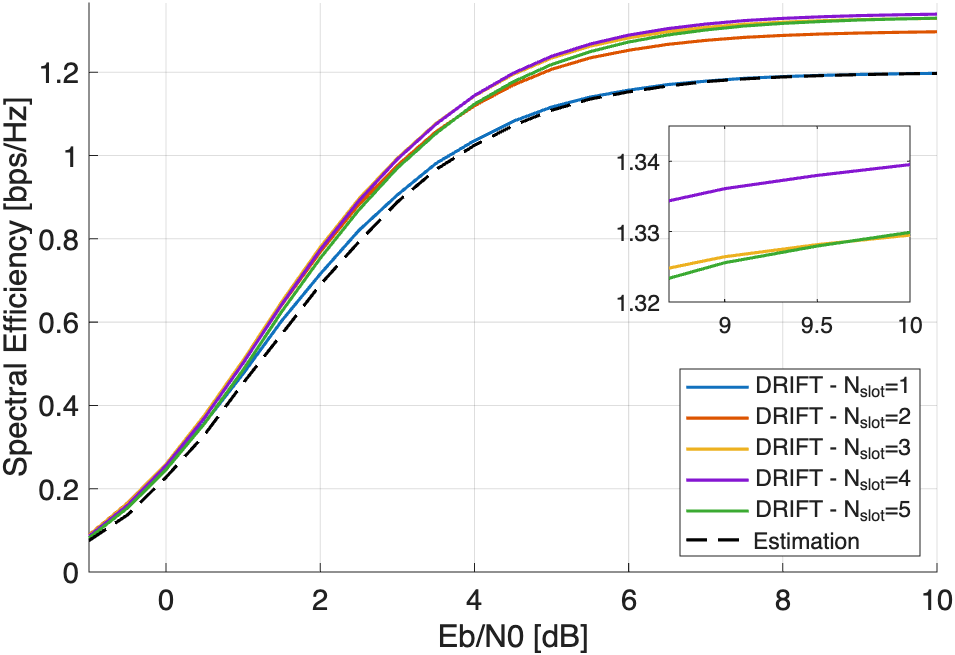}}
    \caption{SE as a function of $E_b/N_0$ for variable $N_{slot}$ (NTN-TDL-C).}
    \label{fig:se_C5}
\end{figure}

\begin{figure}[t]
    \centerline{\includegraphics[width=\columnwidth]{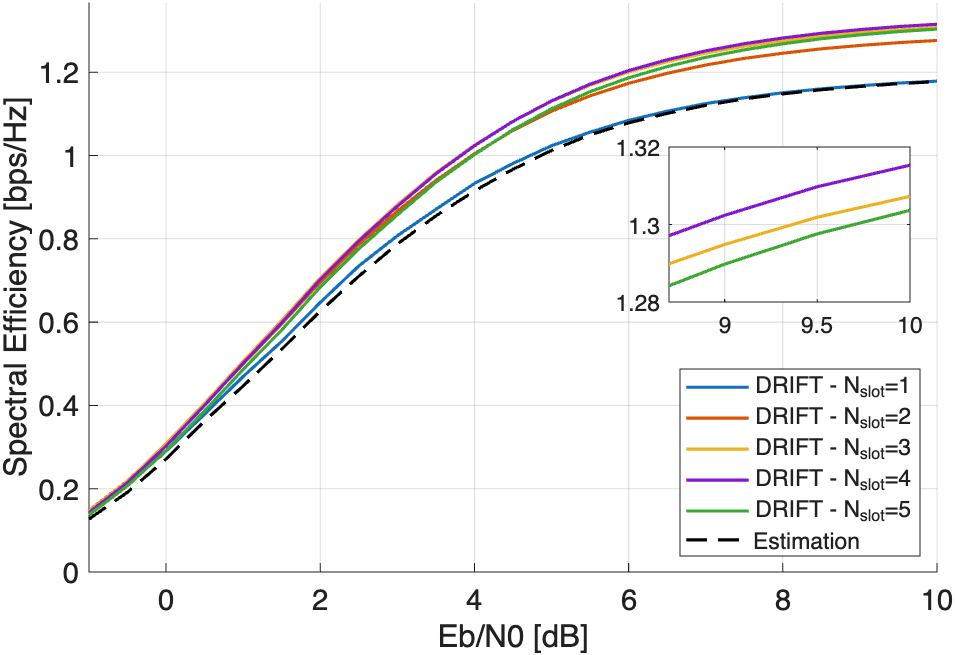}}
    \caption{SE as a function of $E_b/N_0$ for variable $N_{slot}$ (NTN-TDL-D).}
    \label{fig:se_D5}
\end{figure}

We first evaluate the SE gain achieved by the DRIFT models in the proposed framework compared to a traditional estimation-based system (\textit{i.e.}, with pilot-full slots only). The block error rate (BLER) can be assessed as the ratio between the number of erroneous received blocks to the total number of transmitted blocks (hybrid automatic repeat request is assumed to be disabled). Denote with $BLER^{(E,PA)}$ the average BLER in the traditional system and with $BLER_t^{(E')}$ the $t$-th slot's average BLER in the proposed framework. The SE of the estimation-based system can then be evaluated as:
\begin{equation}
    \eta^{(E,PA)} = \frac{m \cdot R_c}{T_{slot} \cdot \Delta f}\cdot (N_{sym}^{slot} - \left| \mathcal{I}_\pi\right|)\cdot(1-BLER^{(E,PA)}),
\end{equation}
where $T_{slot}=1$ ms is the OFDM slot duration according to the select numerology and $m=log_2M$ the number of data bits per symbol. The SE achieved in the proposed system can be evaluated through \eqref{eqn:se_drift}, reported at the top of the next page.
\begin{figure*}[h]
    \normalsize
    \begin{equation}\label{eqn:se_drift}
        \eta^{(DRIFT)}(N_{slot}) = \frac{m \cdot R_c}{T_{slot} \cdot \Delta f}\frac{(N_{sym}^{slot} - \left| \mathcal{I}_\pi\right|)\cdot(1-BLER^{(E')}_1) + N_{sym}^{slot}\sum_{t=2}^{N_{slot}}(1-BLER^{(E')}_t)}{N_{slot}}
    \end{equation}
    \hrulefill
\end{figure*} 
We report in Figure \ref{fig:se_gain} the SE gain at $E_b/N_0=10$dB as a function of $N_{slot}$. The plot shows that both DRIFT models achieve a reduced error propagation compared to the baseline predictors. In particular, the maximum SE gain is reached using the TCN-based DRIFT model, corresponding to a $\approx 12$\% gain over the traditional estimation-based system considering $N_{slot}=4$, \textit{i.e.}, one pilot-full slot and three pilotless slots. As further increasing $N_{slot}$ leads to diminishing returns, this frame structure represents the best tradeoff between SE gain and estimation/prediction accuracy. The Hybrid CNN-LSTM also benefits from the refinement stage introduced by the DRIFT architecture; however, the error propagation reduction is limited by the capabilities of the baseline predictor, and no gain increase is observed after $N_{slot}=3$ in this case. Based on this, we focus on the TCN model for further evaluations. Figures \ref{fig:se_C5} and \ref{fig:se_D5} report the SE achieved by the TCN-based DRIFT model over the NTN-TDL-C and the NTN-TDL-D fading models, respectively, as a function of the $E_b/N_0$ for $N_{slot}\leq5$. The plots show that, up to $N_{slot}=4$, the DRIFT provides a consistent increase in SE at any $E_b/N_0$ working point. Furthermore, setting $N_{slot}=1$ (blue curve) provides a $\approx 0.1$ dB gain in $E_b/N_0$ over the estimation-based system(dashed black curve), confirming that the channel refinement stage produces a more accurate CFR estimate than that available through PA channel estimation. Finally, Figure \ref{fig:se_D5} proves that such findings are also applicable to different NTN-specific LoS fading models without requiring retraining. Indeed, consistent SE gains can be observed up to $N_{slot}=4$ (purple curve), after which the error propagation overcomes the diminishing returns provided by an additional pilotless slot.


\subsection{Normalized mean squared error}\label{ch:4.2_NMSE}
\begin{figure}[t]
    \centerline{\includegraphics[width=\columnwidth]{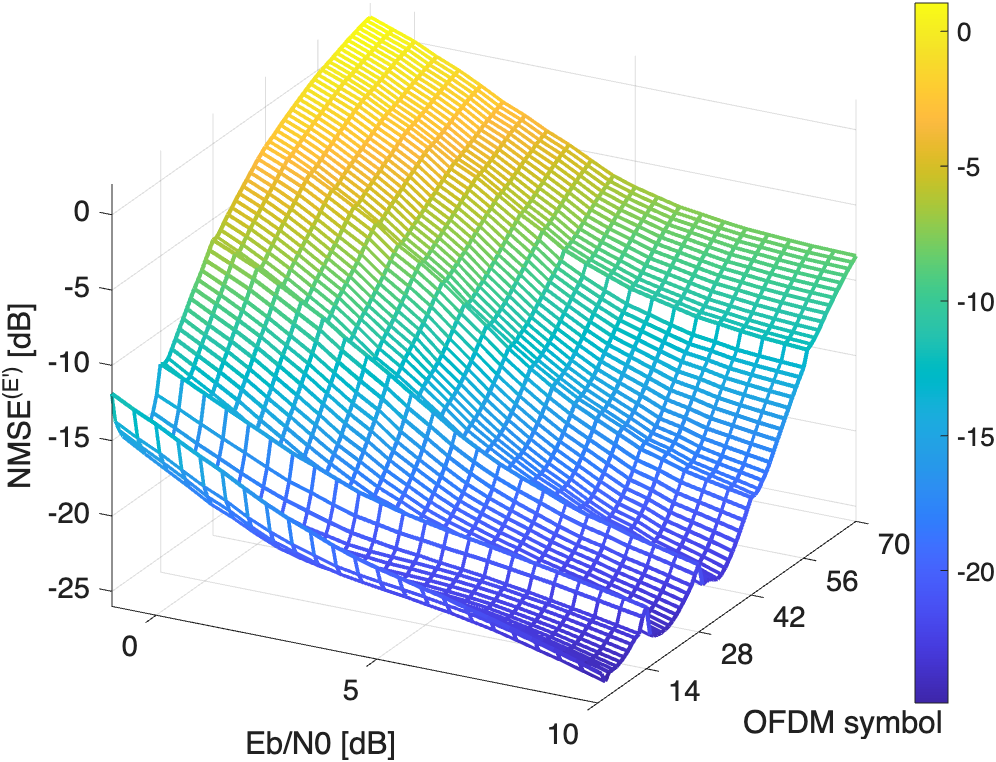}}
    \caption{Estimation NMSE as a function of the $E_b/N_0$ and the OFDM symbol index using DRIFT with TCN (NTN-TDL-C).}
    \label{fig:nmse}
\end{figure}
We report in Figure \ref{fig:nmse} the mesh plot of the channel refinement NMSE averaged over the Monte Carlo tests, where each sample is computed as:
\begin{equation}
    NMSE^{(E')}_{t,k,n} = 
        \frac{\lVert (\hat{\mathbf{H}}_{t}^{(E')})_{k,n} - (\mathbf{H}_{t})_{k,n}\rVert^2}
        {\lVert (\mathbf{H}_{t})_{k,n} \rVert^2}.
\end{equation}
The figure clearly shows that, at high $E_b/N_0$, the refinement NMSE decreases at beginning of each slot until the fifth (starting with the 57th OFDM symbol), where the error propagation leads to an increase in NMSE. Furthermore, the NMSE remains under -19 dB for the entirety of the first three slots, proving the effectiveness of not only the channel refinement section, but also channel prediction.

\subsection{Computational complexity}\label{ch:4.3_complexity}
Finally, we discuss the computational complexity of the proposed DRIFT models, evaluated through the multiply and accumulate units (MACs) metric. We approximate the number of MACs as the number of multiplications in Conv2D, TConv2D, and LSTM layers. A Conv1D layer (including CConv1D layers) with $N_K$ kernels of length $W_K$ having an $L_F \times L_T \times N_C^{(I)}$ tensor as input and an $L_F \times L_T \times N_K$ tensor as output requires the following number of MACs:
\begin{equation}
    MAC^{(Conv1D)} = L_F L_T N_C^{(I)} W_K N_K.
\end{equation}
Based on Figure \ref{fig:CNNscheme}, we conclude that the channel refinement stage results in a total of $\approx 16$k MACs. Taking the figures reported in \cite{bib:JWCN_pred} and \cite{bib:commStd_NOMApred}, the channel prediction stage requires $\approx 138$k MACs and 155k MACs with the Hybrid CNN-LSTM and TCN predictors, respectively. Therefore, the total number of MACs required for inference with the DRIFT models is 154k MACs and 171k MACs for the DRIFT-based Hybrid CNN-LSTM and TCN, respectively, with the channel refinement stage consisting of about 10\% of the overall computational complexity. Such low figures are achieved by design: indeed, channel prediction models tailored to terrestrial networks are not limited by hardware power consumption and can thus invest more MACs to predict CFRs in rich multipath conditions, \textit{e.g.}, the transformer-based P2P model presented in \cite{bib:transformerPilotToPred} requires $\approx 3.5$M MACs (1 MAC = 2 floating point operations) to produce a similar number of elements in the predicted CFR. On the opposite, the DRIFT models are expected to operate on board of a LEO satellite payload, thus being subject to the strict power consumption constraints imposed by the payload's power budget. Nonetheless, recent advances in low-power hardware accelerators have made possible running large CNNs having more than 1M MACs with 100 mW of power at an inference rate of 139 inferences/s, \textit{i.e.}, an inference latency of $\approx 7$ ms \cite{bib:hardware}. Based on this, under the simplified assumption that the inference latency scales with the number of MACs, the Hybrid-CNN-LSTM-based and TCN-based DRIFT models would run at an inference latency of $\approx1.0$ ms/inference and $\approx1.1$ ms/inference, respectively. While such figures are only rough estimates, we note that the authors in \cite{bib:hardware} focused on nano unmanned aerial vehicles and, thus, a higher share of the payload power budget may be allocated to AI accelerators in LEO-based NTNs.

\section{Conclusions}
This paper proposed an iterative joint channel estimation and prediction framework with the objective of reducing the pilot overhead in 6G NTNs. In such framework, a series of $N_{slot}$ slots are transmitted with pilots being carried only in the initial slot. Channel estimation and prediction are based on DRIFT, a data-driven architecture that refines a DA-estimated CFR and predicts the CFR matrix over the upcoming time slot. We presented a lightweight channel refiner based on convolutional layers and implemented two predictors based on convolution and LSTM. The DRIFT models were evaluated in an end-to-end simulation of an uplink transmission towards a LEO satellite. We showed that the TCN-based DRIFT model achieves a SE gain over a traditional estimation-based system of $\approx 12$\% setting $N_{slot}=4$. Furthermore, all evaluated models require under 200k MACs, thus being potentially suitable for real-time on-board inference through state-of-the-art low power hardware accelerators. In the future, we aim at evaluating the real-world inference latency with realistic hardware and to test the DRIFT models with real or emulated data to prove their effectiveness. Furthermore, we aim at further reducing the error propagation effect, \textit{e.g.}, by developing a lightweight pilot-to-estimation-and-prediction model tailored to NTNs. Finally, future research should also focus on assessing and minimizing the impact of non-idealities (\textit{e.g.}, phase noise, multi-user interference) on the estimation and prediction accuracy.

\section{Acknowledgments}\label{Acknowledgment}
This work has been funded by the UNITY-6G project, which received funding from the Smart Networks and Services Joint Undertaking (SNS JU) under the European Union’s Horizon Europe research and innovation programme under Grant Agreement No 101192650. The views expressed are those of the authors and do not necessarily represent the project. The Commission is not liable for any use that may be made of any of the information contained therein.

\bibliographystyle{IEEEtran}
\bibliography{IEEEbib}

@misc{bib:tr38.843,
    author="3GPP",
    title="{TR 38.843 - Study on Artificial Intelligence (AI)/Machine Learning (ML) for NR air interface}",
    year="2024"
}

@article{bib:6GNTN,
    author="Guidotti, A. and Vanelli-Coralli, A. and El Jaafari, M. and Chuberre, N. and Puttonen, J. and Schena, V. and Rinelli, G. and Cioni, S.",
    title="{Role and Evolution of Non-Terrestrial Networks Toward 6G Systems}",
    journal="IEEE Access",
    volume="12",
    pages="55945--55963",
    year="2024",
    note="{doi:10.1109/ACCESS.2024.3389459.}"
}

@article{bib:channelAgingSat,
    author="Abbasi, O. and Kaddoum, G.",
    title="{Channel Aging-Aware LSTM-Based Channel Prediction for Satellite Communications}",
    journal="IEEE Networking Letters",
    volume="6",
    number="3",
    pages="183--187",
    year="2024",
    note="{doi:10.1109/LNET.2024.3444495}"
}

@article{bib:upToDown,
    author="Zhang, Y. and Liu, A. and Li, P. and Jiang, S.",
    title="{Deep Learning (DL)-Based Channel Prediction and Hybrid Beamforming for LEO Satellite Massive MIMO System}",
    journal="IEEE Internet of Things Journal",
    volume="9",
    number="23",
    pages="23705--23715",
    year="2022",
    note="{doi:10.1109/JIOT.2022.3190412}"
}

@article{bib:SCP,
    author="Zhang, Y. and Wu, Y. and Liu, A. and Xia, X. and Pan, T. and Liu, X.",
    title="{Deep Learning-Based Channel Prediction for LEO Satellite Massive MIMO Communication System}",
    journal="IEEE Wireless Communications Letters",
    volume="10",
    number="8",
    pages="1835--1839",
    year="2021",
    note="{doi:10.1109/LWC.2021.3083267}"
}

@INPROCEEDINGS{bib:attentionPred,
  author={Cui, Chuankai and Jing, Wenpeng and Lu, Zhaoming and Wen, Xiangming},
  booktitle={2024 IEEE 35th International Symposium on Personal, Indoor and Mobile Radio Communications (PIMRC)}, 
  title="{Attention Aided Channel Prediction Scheme For Satellite-Terrestrial Networks}", 
  year={2024},
  volume={},
  number={},
  pages={1-5},
  doi={10.1109/PIMRC59610.2024.10817356}}

@article{bib:commStd_NOMApred,
  author={Amatetti, Carla and Filippo, Bruno De and Vanelli-Coralli, Alessandro},
  journal={IEEE Communications Standards Magazine}, 
  title="{Channel Prediction With Temporal Convolutional Networks: A New Paradigm to Enable Non-Orthogonal Multiple Access in 6G NTN}", 
  year={2026},
  volume={},
  number={},
  pages={1-9},
  doi={10.1109/MCOMSTD.2026.3666960}
}

@article{bib:JWCN_pred,
  author = "De Filippo, Bruno and Amatetti, Carla and Vanelli-Coralli, Alessandro",
  title = "{In-orbit channel prediction: deep learning architectures for 6G non-terrestrial networks}",
  journal = "Journal on Wireless Communications and Networking",
  year = "2026",
  volume={},
  number={},
  doi = {10.1186/s13638-026-02604-x},
  note = "In print."
}

@misc{bib:tr38.811,
    author="3GPP",
    title="{TR 38.811 - Study on New Radio (NR) to support non-terrestrial networks (Release 15)}",
    year="2020"
}

@article{bib:sync,
    author="Lin, J.-C.",
    title="{Synchronization Requirements for 5G: An Overview of Standards and Specifications for Cellular Networks}",
    journal="IEEE Vehicular Technology Magazine",
    volume="13",
    number="3",
    pages="91--99",
    year="2018",
    note="{doi:10.1109/MVT.2018.2813339}"
}

@ARTICLE{bib:hardware,
  author={Lamberti, Lorenzo and Bellone, Lorenzo and Macan, Luka and Natalizio, Enrico and Conti, Francesco and Palossi, Daniele and Benini, Luca},
  journal={IEEE Internet of Things Journal}, 
  title="{Distilling Tiny and Ultrafast Deep Neural Networks for Autonomous Navigation on Nano-UAVs}", 
  year={2024},
  volume={11},
  number={20},
  pages={33269-33281},
  doi={10.1109/JIOT.2024.3431913}
}

@ARTICLE{bib:transformerPilotToPred,
  author={Lagona, Louis and Vakilifard, MohammadAmin and Bockelmann, Carsten and Dekorsy, Armin},
  journal={IEEE Wireless Communications Letters}, 
  title="{Transformer-Based Pilot-to-Prediction for Frequency-Selective Channels in OFDM Systems}", 
  year={2026},
  volume={15},
  number={},
  pages={310-314},
  doi={10.1109/LWC.2025.3625969}
}

@misc{bib:aNB,
  title        = {{TDoc R3-260002 - RAN3 \#131 Meeting Report}},
  author       = {{3GPP MCC}},
  institution  = {{3rd Generation Partnership Project (3GPP)}},
  address      = {Dallas, Texas, United States},
  year         = {2025}
}

@ARTICLE{bib:TFformer_pred,
  author={Yan, Daifu and Jia, Min and Guo, Qing and Niyato, Dusit},
  journal={IEEE Transactions on Wireless Communications}, 
  title="{Temporal-Frequency Domain Channel Prediction for LEO Satellite Communication System: A Novel TFformer Structure}", 
  year={2025},
  volume={24},
  number={10},
  pages={8208-8220},
  doi={10.1109/TWC.2025.3564713}
}

@ARTICLE{bib:QV_pred,
  author={Homssi, Bassel Al and Chan, Chiu C. and Wang, Ke and Rowe, Wayne and Allen, Ben and Moores, Ben and Csurgai-Horváth, László and Fontán, Fernando Pérez and Kandeepan, Sithamparanathan and Al-Hourani, Akram},
  journal={IEEE Transactions on Machine Learning in Communications and Networking}, 
  title={Deep Learning Forecasting and Statistical Modeling for Q/V-Band LEO Satellite Channels}, 
  year={2023},
  volume={1},
  number={},
  pages={78-89},
  doi={10.1109/TMLCN.2023.3286793}
}

\end{document}